\documentclass[preprint]{sig-alternate-05-2015}
\PassOptionsToPackage{hyphens}{url}\usepackage{hyperref}
\usepackage{breakurl}
\usepackage{flushend}
\def\draftmode{}
\usepackage{pslatex}
\usepackage{subfigure}
\usepackage[usenames]{color}

 \ifx\draftmode\defined
  \newcommand{\todobox}[1]{}%
  \else
 \newcommand{\todobox}[1]{%
     \centering
    \fbox{\parbox[l]{\columnwidth}{\textcolor{blue}{TODO: #1}}}\\%
 }
 \fi 
 
 \ifx\draftmode\defined
  \newcommand{\myreason}[1]{}%
  \else
 \newcommand{\myreason}[1]{%
     \centering
    \fbox{\parbox[l]{\columnwidth}{\textcolor{green}{REASONING: #1}}}\\%
 }
 \fi 
 
\ifx\draftmode\undefined
\newcommand{\mycomment}[1]{}%
\else
\newcommand{\mycomment}[1]{{\textcolor{red}{ \bf $<$#1$>$}}}
\fi 

\begin{document}

%\setcopyright{acmcopyright}
%\conferenceinfo{NS Ethics'15,}{August 17-21, 2015, London, United Kingdom}
%\isbn{978-1-4503-3541-6/15/08}\acmPrice{\$15.00}
%\doi{http://dx.doi.org/10.1145/2793013.2793023}
\toappear{}

\title{Contextual Consent: Ethical Mining of \\
	Social Media for Health Research}

\numberofauthors{2}
\author{
	\alignauthor Chris Norval\\
		\affaddr{School of Computer Science}\\
	    \affaddr{University of St Andrews}\\
	    \affaddr{St Andrews, Fife, UK}\\
	    \email{cn56@st-andrews.ac.uk}    
	\alignauthor Tristan Henderson\\
	    \affaddr{School of Computer Science}\\
	    \affaddr{University of St Andrews}\\
	    \affaddr{St Andrews, Fife, UK}\\      
	    \email{tnhh@st-andrews.ac.uk}
}

\sloppy
\hyphenpenalty 10000

\makeatletter
 \hypersetup{%
  breaklinks,
  baseurl       = http://,%
  pdfborder     = 0 0 0,%
  pdfpagemode   = UseNone,%
  colorlinks    = false,%
  pdfstartpage  = 1,%
  pdfcreator    = \LaTeX{},%
  pdfproducer   = \LaTeX{},%
  pdfauthor     = {},
  pdftitle      = \@title,
  pdfkeywords   = {ethics, health, social media}
  }
\makeatother

\maketitle

\begin{abstract}
Social media are a rich source of insight for data mining and
user-centred research, but the question of consent arises when
studying such data without the express knowledge of the creator. Case
studies that mine social data from users of online services such as
Facebook and Twitter are becoming increasingly common.  This has led
to calls for an open discussion into how researchers can best use
these vast resources to make innovative findings while still
respecting fundamental ethical principles. In this position paper we
highlight some key considerations for this topic and argue that
the conditions of informed consent are often not being met, and that
using social media data that some deem free to access and analyse may
result in undesirable consequences, particularly within the domain of
health research and other sensitive topics. We posit that successful
exploitation of online personal data, particularly for health and
other sensitive research, requires new and usable methods of obtaining
consent from the user.

\end{abstract}


\section{Introduction}
\label{s:intro}

Social media services, such as Facebook and Twitter, are a near
ubiquitous part of people's lives. Facebook, for example, boasts over
1.7 billion monthly active users~\cite{facebook:newsroom}. As a result 
we are increasingly sharing more and more data about our lives online: 
68\% of UK-based online adults look at social media sites or apps each 
week, 35\% upload photos or videos, 28\% share links to websites or 
online articles, and 24\% contribute comments to a website or
blog~\cite{ofcom-attitudes:2016}. %p74

%[active users], who use it for a variety of reasons including
%communication with close friends, self-presentation, fostering support 
%networks, relieving boredom, meeting new people, and finding 
%information aboutothers~\cite{blachino:2013}

The vast quantities of personal data shared through such platforms can
offer new insights into important areas of research~\cite{boyd:2012}.
Academics from various fields have capitalised on these data, leading
to the comprehension of complex social behaviours and trends in
society. Within a health context alone, such social data has been
found to act as a predictor for
depression~\cite{moreno:2011,reece:instagram}, suicide risk
factors~\cite{choudhury:2016,jashinsky:2014}, mood
changes~\cite{lee:2016}, flu outbreaks~\cite{li:2013} and problem
drinking in US college students~\cite{moreno:2012}. Such findings can
help researchers and practitioners identify important markers in
health and in society.

Despite these advantages, those who created the social data in
question are not always informed about the process. While some
researchers outline clear methods of obtaining informed consent when
conducting research on social media data (e.g.  \cite{moreno:2012}),
only around 5\% of the literature that use such data
describe their consent procedures \cite{hutton:reproducibility}.

Some have argued that social data posted online 
are freely available for research~\cite{hudson:2004}; however, others 
argue that just because personal information is made available online 
does not mean it is appropriate to capture and analyse~\cite{boyd:2012, 
conway:2016, hutton:consent, zimmer:2010}. Conway and O'Connor 
argue that the potential challenge to privacy occurs \textit{``not in the 
reading or accessing of individual materials (publicly available as they 
are), but rather in the processing and dissemination of those materials in 
a way unintended"}~\cite{conway:2016}.

While it is important for such research using social media data to
continue, the informed consent of all participants is an important
milestone to strive for. Firstly, consent decisions are strongly tied
to the expected audience~\cite{luger:informed,mcneilly:concerns}.
Research has suggested that users have differing online behaviour
based on their perceived audience, including more
self-censorship~\cite{das:2013,sleeper:2013}, and sharing smaller
proportions of both positive and negative emotions~\cite{burke:2016}
when the perceived audience is more sparse, and less controlled. 

Secondly, research into privacy settings on social media have 
suggested that most users may be significantly oversharing what is 
intended~\cite{liu:2011,madejski:errors}. When analysing social media 
data online, data can be taken from individuals who are unaware that it 
is accessible, and privacy violations may be occurring. The likelihood of 
privacy violations increases with the number of people involved, with
many such studies indiscriminately scraping the data of huge numbers 
of individuals~\cite{boyd:2012}, or not even reporting how they gather
data~\cite{hutton:reproducibility}. 

And finally, participants may find it intrusive to discover that their
data is used in a way which is outside the boundaries of what was
originally expected~\cite{conway:2016, luger:informed}. An example of
this is Samaritans Radar, a well-intentioned Twitter app designed to
monitor the tweets of the user's contacts for potentially suicidal
messages~\cite{lee:radar}. This app received criticism from the
Twitter community over privacy concerns, and was shut down days 
after launching.

Therefore, with all previous points considered, participants may be 
sharing more sensitive and emotional content with a far wider audience 
than intended, they may not have otherwise provided consent for their 
data to be studied, and would likely find a violation of this to be  
intrusive. This suggests a disconnect between the researcher and the 
participant; a situation in which the autonomy of the participant should 
be respected. Safer and more explicit methods of consent are therefore 
necessary to best protect the needs of potential participants, while 
ensuring that we can make the best use of the vast quantities of social 
media data for health research.

In this paper, we take the position that new forms of obtaining
meaningful consent are necessary for successful online health
research. We outline current methods of obtaining 
consent, discuss new potential methods that seek to mitigate these 
problems, and outline some of the challenges that need 
to be overcome for such a method to be developed.

\section{Background} 
\label{s:background}

Informed consent, a declaration that the participant understands the 
consequences of participating in the study, is widely seen as 
fundamental to conducting research with human participants
\cite{manson:consent}. From the point of view of the researcher, it is 
seen as fulfilling ethical responsibilities with regards to the protection of 
data, privacy, and autonomy of the 
participant~\cite{morrison:personalised-representations}. While there 
appears to be broad agreement when it comes to interventional 
research, the subject of whether informed consent is necessary in a 
study of online communications is the subject of 
debate~\cite{hudson:2004,solberg:2010}. As a result, new models 
for consent have been proposed~\cite{hutton:consent,kaye:dynamic}.

\subsection{User attitudes and the need for consent}
 Studies into user attitudes are often subject to selection bias, given 
 that  participants have opted in and are therefore not necessarily 
 representative of the wider population. Despite this, it is worth 
 consulting the viewpoints of those who may be potentially affected by 
 such research, and how they approach the trade-off between societal 
 benefit and personal intrusion. Mikal et al. investigated user attitudes 
 toward the analysis of social media data for research, discovering 
 \textit{``equivocal findings"} in the 
 literature~\cite{mikal:depression-monitoring}. 

Mikal et al. asked participants about their expectations of privacy with 
regard to monitoring depression at the population level, finding that  
most were accepting of it~\cite{mikal:depression-monitoring}. This was, 
however, largely conditional on the analysis being conducted in an 
aggregated and anonymised way, with no way of targeting a particular
individual. This is echoed by Conway and O'Connor, who argue the 
need for a differentiation between automatic identification at the 
individual level and at the population level in order to protect the privacy 
of those involved~\cite{conway:2016}. 

Mikal et al. found that \textit{``many respondents felt as though a 
failure to protect online data constituted consent to have that data 
systematized and analyzed"}, however, the authors go on to consider 
what kinds of individuals would be less likely to protect their data 
\cite{mikal:depression-monitoring}. Such a model would leave those 
with limited Internet literacy skills or those mistakenly oversharing their 
social data at risk of unwittingly providing implied consent.

\subsection{Unsuitability of informed consent}
Research has questioned the suitability of informed consent as it
currently 
stands~\cite{hutton:consent,kaye:dynamic,luger:informed,
	morrison:personalised-representations,munteanu:situational}.
 Luger and Rodden argue that consent, as currently conceived, cannot 
hope to meet the challenges posed by ubiquitous computing systems, 
and tensions exist between what is expected of consent and what is 
expected of a ubiquitous computer system. They describe consent as 
having been \textit{``stretched thin to the point of 
breaking"}~\cite{luger:informed}.

Kaye et al. argue that current proposals for the European Data 
Protection Regulations\footnote{since the writing of that paper, the 
GDPR has been ratified and will come into effect in 2018.} require 
explicit consent, questioning the legality of broad consent methods, 
and that static systems for obtaining consent are no longer fit for 
purpose due to the changing nature of research and technology. New 
approaches are therefore needed to meet the ethical and legal 
requirements for consent while accomodating the dynamic nature of 
modern research~\cite{kaye:dynamic}.

Some of these concerns have been echoed by Morrison et al., who 
found that few participants were aware of participating in an academic 
trial, as specified in the Terms and Conditions document of a mobile 
application. Further means of informing users of researchers' intentions 
to collect and analyse user data are necessary to behave in an ethical 
manner~\cite{morrison:personalised-representations}. Such findings 
call into question the suitability of broad, one-off methods of collecting 
consent when dealing with online research. As Steinsbekk et al. 
articulate, at the core of the debate is \textit{``what it means to be 
`adequately informed' and whether giving consent based on broader 
premises is valid or not"}~\cite{steinsbekk:dynamic}.

\subsection{Dynamic consent}
Researchers have investigated different mechanisms for coping with 
the above issues, such as using dynamic consent~\cite{kaye:dynamic,
	munteanu:situational,steinsbekk:dynamic}. Dynamic consent is a 
framework for allowing participants to grant access to their personal 
data to researchers in a way that they can control. For example, a 
participant may decide to revoke access to their data, or customise their 
opt-in/opt-out preferences for participating in research. 

Kaye et al. argue that, for participants, some of the advantages of 
dynamic consent include the ability to easily consent to new projects, 
alter consent preferences in real time, find out how their data has 
been used, and to set preferences about how they are kept informed. 
Researchers are said to gain more engaged participants, streamlined 
recruitment, improved public trust, and the knowledge that their  
research conforms to high legal standards~\cite{kaye:dynamic}. 

While researchers have outlined the potential benefits of dynamic 
consent, Steinsbekk et al. dispute these, arguing that a convincing case 
has not been made. Their criticisms of dynamic consent include more 
frequent (and therefore more trivial) requests for re-consent, and an 
increased risk of the relationship between researcher and participant 
breaking down due to unmet expectations and a lack of 
reciprocity~\cite{steinsbekk:dynamic}. The authors argue that 
\textit{``broad consent combined with competent ethics review and an 
active information strategy is a more sustainable solution"}. Hutton and 
Henderson have also raised issues with dynamic consent, arguing that 
frequent requests for consent can lead to a significantly greater burden 
on the participant~\cite{hutton:consent}, which could frustrate 
participants and lead them to withdraw from the research.

\subsection{Contextual integrity}
Nissenbaum's model of contextual integrity is a theoretical framework 
that attempts to prescribe \textit{``specific restrictions on collection, 
use, and dissemination of information about people''} depending on 
presiding norms of information appropriateness and 
distribution~\cite{nissenbaum:integrity}. Contextual integrity focuses on 
whether a flow of information is appropriate within a particular context, 
or whether a violation of privacy has occurred~\cite{sar:2013}. It has 
previously been used to explore privacy implications of social networking 
sites~\cite{barkhuus:2012,hull:2011,hutton:consent,sar:2013,zimmer:2010}.

Hutton and Henderson have used contextual integrity to explore a new 
method of obtaining consent for social media research, described as 
\textit{`contextual integrity consent'}~\cite{hutton:consent}. This 
middle-ground approach seeks the flexibility of dynamic 
consent, allowing users to choose what data is accessible and when, 
while reducing the burden of such data management. It works by 
inferring context-specific norms; a ruleset determining the appropriate 
flow of information. Individuals are asked explicitly about their willingness 
to share unless they clearly conform to or deviate from such a norm. In 
situations where a violation is found to have occurred, the framework 
rejects the practice in question and further actions can be taken, such 
as re-requesting consent from the user.

These norms depend on a number of factors, such as what the data are,
to whom the data are flowing, and for what purpose they are being
requested. For example, McNeilly et al. found that participants of a 
location sharing study were more likely to give the researchers access 
to their location data than a health study, and in both studies, photos 
were much less likely to be shared than `liked' 
pages~\cite{mcneilly:concerns}. These norms can change over time, 
and individuals, organisations, and sections of society can each have 
their own expectations of what is appropriate. 

%For example, it may be acceptable to share holiday photos on 
%social media with friends and family, but not with medical professionals. 
%Messages shared with medical support communities, on the other hand, 
%may be appropriate for medical professionals and researchers, but not 
%colleagues. 

%Such a flow of data can be risky when, for 
%example, the rules and norms are not well established, understood, or
%respected~\cite{lee:radar}. 

%\newpage
\section{Towards contextual consent for health}
\label{s:project}

Much of the work in dynamic consent has studied mobile or social
network applications. An open question is how to apply these
techniques to mining online health data. Our aim is to investigate
contextual consent for health; can we capture, interpret, and act on
context-specific norms within a healthcare domain? Further, can we
improve on contextual consent by blending contextual integrity with 
machine learning techniques? Finally, such techniques are only useful if 
available to practitioners, and feedback from such deployments will 
inform further development of tools and models.

By predicting when social media users find it appropriate to share 
different types of data with different stakeholders, such as researchers 
and clinicians, the consent process will, firstly, better reflect the context 
in which data were created, and secondly, respect users' preferences 
about which data should be made available and with whom. For 
instance, someone seeking support from their peers because they are 
anxious about an upcoming medical procedure might not want this 
shared with medical researchers, while others may want reports about 
side-effects of their medication to be viewed by clinicians. 

We are focusing specifically within a health context for three reasons. 
Firstly, we believe that health data is an area where presiding norms 
exist. Secondly, getting these expectations of appropriate data flow 
correct is important due to the potential sensitivity of the data. 
Finally, given the vast number of studies using social media data within 
a medical domain, we believe this work is timely, and there is an 
opportunity to provide guidance and tools for the appropriate handling 
of social media data in health research to academics and practitioners 
alike. 

We are currently planning two studies to further this investigation. The 
first will involve collecting a large corpus of data to use in a training and 
model evaluation phase. We will work with participants to determine 
what the main predictors are of appropriate data flow to different 
medical stakeholders. With this data, we will then derive 
machine-learning models for predicting consent with these different 
stakeholders. Finally, we will evaluate our model's effectiveness in a 
follow-up study with participants in a social media medical support 
community.

Gomer et al. propose a similar semi-autonomous method of obtaining
consent. Their proposed system trains a consent agent, uses this 
model to receive and accept or reject requests for participation, and 
allows users to review past decisions with the results re-training the
model~\cite{gomer:agents}. This method differs from ours as it trains 
and builds a model on a person-by-person basis, rather than building 
on the collective norms which shape contextual integrity. As such, it 
would require user training before it could be used. At the time of 
writing, no follow-up work has since been published.

Throughout this research, we aim to engage with clinicians, members of 
support communities and researchers to understand the concerns and 
interests of all parties. Due to the potential for sensitive topics and the 
extent in which a sharing violation may impact individuals, it is important 
that steps are put in place to greatly reduce risk. Such a predictive 
model will likely serve as a recommender system until the technique can 
be proven, the risks are understood, and confidence in such a method 
is developed. Working closely with all relevant stakeholders will help us 
to document and understand the risks involved and the challenges that 
will need to be overcome.

%\newpage
\section{Challenges}
\label{s:challenges}

One of the largest technical challenges we face is the contextual nature 
of consent over items of data; One shared status update or photo may 
be seen as appropriate for researchers to access and utilise, whereas 
another may not be, making it difficult to design broad rules for what is 
acceptable and what is not \cite{luger:informed, madejski:errors}. 
Furthermore, the given rules for any one individual may change over 
time, as social relationships and opinions evolve \cite{liu:2011}.

Participants may have concerns about sharing sensitive data with any 
autonomous system, given that it is an extension of the researcher. This 
raises new challenges about how such a classifier could be used, or 
even trained, if some participants are not willing to share certain data 
with researchers. Approaching this project, we must strike a balance 
between accuracy and appropriateness, as participants may or may not 
be comfortable with having the actual content of their social data being 
mined, however, it is unclear how effective prediction techniques can be  
using metadata alone. 

A significant ethical consideration is the potential for erroneous and 
unacceptable norm violations taking place as a result of this research. 
In short, the sharing of a particular bit of data against the will of the 
owner may erode trust and call into question the entire project. Such a 
system may not be given a second chance. Our interim solution to this 
is to focus on a system which suggests appropriate sharing levels to 
the user for approval, rather than explicitly sharing the data 
autonomously. While this increases the burden of sharing content, it 
would still be less burdensome than asking the user to specifically 
choose the acceptable audience on each bit of content shared, such as 
with groups, lists, or circles. We believe that this limitation exists in any 
system that aims to reduce the burden of consent by automating 
the user's decision.

The collection of data for such research will also likely prove to be a 
barrier to overcome. Collecting any data in an ethically aware way will 
introduce selection bias. Obtaining large sample sizes may also be 
difficult. We believe that this challenge is also inherent in such 
research, although there may be ways to minimise the impact without 
compromising the principle of informed consent.

%\newpage
\section{Conclusion}
\label{s:conc}

People are increasingly sharing more and more information via social 
media, and as a result, such platforms are seen as rich data sources for 
researchers from various fields and backgrounds. This can lead to 
interesting research and discoveries, however, it also raises significant 
questions over informed consent in the age of social media and data 
mining. We argue that the successful exploitation of online personal 
health data requires new and usable methods of obtaining consent from 
the content creators. By deriving a new and usable method of obtaining 
consent in such circumstances, we hope that researchers can continue 
safe in the knowledge that the research is transparent and the 
participants are informed about how their data is being used and why.

We are currently planning studies to explore the contextual norms of 
data sharing within the context of health research in order to investigate 
whether or not this process can be automated. To develop a new 
method of obtaining consent, we need a broad range of input in the 
areas of social media, privacy, ethics, and law. We would like to invite 
experts, researchers and practitioners  in these fields, as well as users 
of such services, to provide thoughts and contribute to discussions 
about how such methods may work.

\section*{Acknowledgements}
This work was supported by the Wellcome Trust [UNS19427].

% https://wellcome.ac.uk/funding/managing-grant/research-publication-acknowledgement-practice-guidance-authors
 % not in the double-blind submission

%\bibliographystyle{abbrvDOI}
\bibliographystyle{abbrv}
\bibliography{mohrs2017}

\balancecolumns
\end{document}